\documentclass[twocolumn,pra,showpacs,preprintnumbers]{revtex4}
\usepackage{amsmath,amssymb}
\usepackage{graphicx}
\usepackage{dcolumn}
\usepackage{bm}
\newcommand{\erf}{\mathrm{erf}}
\newcommand{\sgn}{\mathrm{sgn}}

\newcommand{\ee}{\mathrm{e}}
\newcommand{\ii}{\mathrm{i}}

\newcommand{\sig}{\sigma}

\newcommand{\lL}{{\mathcal{L}}}

\newcommand{\be}{{\mu}}

\begin{document}
\author{S. F. Caballero Ben\'\i tez$^{1,2}$, E. A. Ostrovskaya$^1$, M. Gul\'acs\'\i$^{1,2}$ and Yu. S. Kivshar$^1$} 
\affiliation{$^{1}$ARC Centre of Excellence for Quantum-Atom Optics and Nonlinear Physics Centre, Research School of Physical Sciences and Engineering, Australian National 
University, Canberra ACT 0200, Australia}
\affiliation{
$^{2}$ 
Max-Planck-Institut f\"ur Physik Komplexer Systeme, 
N\"othnitzer Strasse 38,
01187 Dresden
Germany}

\title{Macroscopic quantum self-trapping in a Bose-Josephson junction with fermions.}
\begin{abstract}
We study the macroscopic quantum self-trapping effect in a mixture of Bose-Einstein condensate and a large number of quantum degenerate fermions, trapped in a double-well potential with a variable separation between the wells. The large number of fermions localized in each well form quasi-static impurities that affect the dynamics of the bosonic cloud. Our semi-analytical analysis based on a mean-field model shows that main features of macroscopic  quantum self trapping in a pure bosonic system are radically modified by the influence of fermions, with both the onset of self-trapping and properties of the self-trapped state depending on the fermion concentration as well as on the type of inter-species interaction. Remarkably, repulsive inter-species interaction leads to population inversion of the bosonic energy levels in the trapping potential and hence to the inversion of symmetry of the macroscopic wavefunction. Realistic physical estimates are given based on experimental parameters for a $^{40}$K-$^{87}$Rb system. 
\end{abstract}
\maketitle
\section{Introduction}
Self-trapping phenomena are  among the most dramatic effects of atomic interactions \cite{RevModPhys.73.307} in the systems of quantum degenerate gases. The so-called \emph{macroscopic quantum self-trapping}  (MQST) effect \cite{PhysRevLett.79.4950} manifests itself as localization of most of the particles in the system in a particular region in space. The MQST and related effects in effectively purely bosonic systems have been extensively analyzed in different physical contexts, from the Josephson effect in superconductors \cite{PL-1-7-251} and the study of surperfluid He$^4$ \cite{RevModPhys.38.298} to the alkali Bose-Einstein condensates \cite{bjj}.  

\begin{figure}[ htbp!] 
   \centering
   \includegraphics[width=7cm,keepaspectratio]{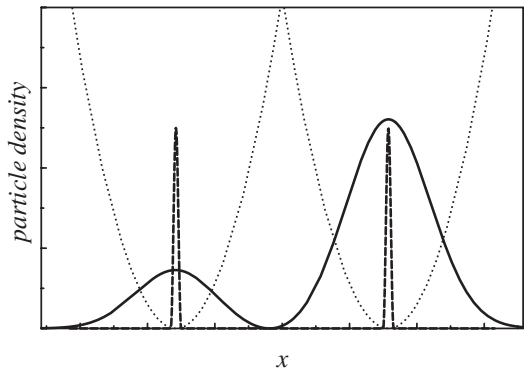} 
   \caption{Schematics of the atomic density distribution for fermionic impurities (dashed line) and bosonic condensate (solid) in a macroscopic self-trapped stationary state formed in a one-dimensional double-well trapping potential (dotted).}
   \label{schematics}
\end{figure}

The MQST effect in a so-called Bose Josephson junction, i.e. a  Bose-Einstein condensate loaded into a double-well potential, has been extensively studied theoretically \cite{PhysRevA.61.031601, PhysRevA.59.620} and observed experimentally \cite{albiez:010402}. Its appearance is linked to a nonlinearity-induced phase-locking between the nonlinear eigenstates of the system, which leads to the formation of new stationary states in the trapping potential that have no linear counterpart \cite{PhysRevA.61.031601,dagosta}. These states are characterized by a population imbalance between the wells of the trapping potential, which becomes more pronounced with growing nonlinearity (see Fig. \ref{schematics}). Moreover, it turns out that the MQST effect plays an important role in the dynamics of condensates in periodic potentials, leading to the formation of self-trapped or truncated gap states \cite{st_markus,st_tristram}.

The experimental achievements in the production of mixtures of  BECs and degenerate Fermi gases with inter-species interaction strength controlled by Feshbach resonances 
\cite{ferlaino:040702, PhysRevLett.87.080403, PhysRevLett.89.150403, stan:143001}, 
drew attention to the studies of the Bose-Fermi  systems in the context of novel quantum phases and lattice dynamics. Properties of the nonlinear spatially localized states, Bose-Fermi solitons, have also been studied both in harmonic \cite{bf_trains,bf_trains2,bf_soliton} and periodic trapping potentials \cite{salerno:063602}. In contrast, this paper presents a theory of the self-trapping effect in the basic model of a Bose-Fermi mixture loaded into a double-well potential. As  demonstrated by the previous research into purely bosonic systems \cite{st_markus,st_tristram}, such an effect will also have consequences for the behaviour of the quantum degenerate mixture in periodic potentials.

Our aim here is to describe the effect of degenerate fermions on the self-trapping behavior of ultracold bosons in a quasi-one dimensional symmetric double-well potential \cite{albiez:010402}, in the limit of a large number of fermions \cite{ospelkaus:020401}. In this limit, the ultracold fermions are not likely to engage in a collective motion and, in a presence of a confining potential, form a quasi-static inhomogeneous distribution of localized impurities in the bosonic cloud (see Fig. \ref{schematics}). In the case of a double-well potential we develop analytical treatment of the system, based on a quasi-static density approximation for the fermionic component and a coupled-mode theory for the bosonic component. We consider both attractive and repulsive interactions between bosons and fermions, which can be modified by means of  Feshbach resonances \cite{ferlaino:040702, ospelkaus:020401, silber:170408, albiez:010402, stan:143001,  PhysRevLett.89.150403, PhysRevLett.87.080403}, in particular considering the system of ultracold $^{40}$K-$^{87}$Rb atoms.  We analyze the effect of a quasi-static distribution of fermions on the formation of MQST states in the interacting bosonic component of the system and show the difference of the physics between the attraction or repulsion between fermions and bosons. The two types of interaction between the atomic species can be achieved experimentally \cite{ferlaino:040702}, which offers a unique opportunity to explore different MQST phases in the system.

The paper is organised as follows: Section II introduces the second quantized and the effective one-dimensional mean-field model for the quantum degenerate Bose-Fermi mixture trapped in a one-dimensional double-well potential. Section III presents our basic approximation of the quasi-static density distribution for fermions and  derivation of the coupled mode theory for the macroscopic BEC wavefunction. Section IV presents the analysis of the self-trapping regimes based on the coupled-mode theory and variational ansatz for the BEC wavefunction in the two main cases of the inter-species attraction and repulsion. Section V concludes and summarizes our work. The details of the variational calculations are given in the Appendix. 

\section{The mean-field model}
We consider a Bose-Fermi mixture trapped in a one dimensional symmetric double-well potential. The Hamiltonian of the system  can be written in second quantized form in terms of single particle operators for each of the particle species and the inter-species interaction between bosons and fermions:
\begin{eqnarray*} \label{ham}
\hat{\mathcal{H}}=
\int_{\mathbb{R}} \mathrm{d}x
\left\{\sum_{\xi\in\{ f,b\}}\right.
\hat{\psi}^\dagger_{\xi}  \hat{H}^{\phantom{\dagger}}_\xi\hat{\psi}^{\phantom{\dagger}}_{\xi}  
&+&\frac{g_{bb}}{2}
 \hat{\psi}^\dagger_{b}  \hat{\psi}^\dagger_{b}  
\hat{\psi}^{\phantom{\dagger}}_{b}  \hat{\psi}^{\phantom{\dagger}}_{b}  
\\
&+&\left.g_{bf}  \hat{\psi}^\dagger_{b}  \hat{\psi}^\dagger_{f}  
\hat{\psi}^{\phantom{\dagger}}_{b}  \hat{\psi}^{\phantom{\dagger}}_{f}  
\right\},
\end{eqnarray*}
where the bosons and fermions in the system are represented by their corresponding field operators: $\hat{\psi}^\dagger_{b,f}$  (creation) and $\hat{\psi}^{\phantom{\dagger}}_{b,f}$ (annihilation); $  \hat{\psi}_\xi\equiv\hat{\psi}_\xi({\bf r},t)$. The field operators  obey the usual commutation (bosons) and anti-commutation (fermions) algebra.

The longitudinal ($x$-component) single-particle part of the Hamiltonian (\ref{ham}) includes a one-dimensional double-well trapping potential:
$
\hat{H}_{\xi x}={\hat{p}_{x}^2}/({2 m^{\phantom{2}}_\xi})+{m^{\phantom{2}}_{\xi}\omega_{\xi}^2}(|x|-x_0)^2/{2}
$.
where the trapping frequencies and the masses for each of the species are denoted by $\omega_{b,f}$ and $m_{b,f}$, respectively, and the distance between the minima of the double-well is given by $2 x_0$.

The interaction strengths $g_{bb}$ and  $g_{bf}$ in Eq. (\ref{ham}) represent the scattering between bosons and boson-fermion, respectively. These coefficients are considered to be constant in the relevant region near their resonances and approximately proportional to their s-wave scattering length.  The on-site fermion-fermion interaction is Pauli suppressed. In the following sections we discuss both he inter-species repulsion and attraction, and show that the sign of the interaction leads to very different physics.

Following the treatment given in \cite{salerno:063602}, we  derive the equations of motions from the Hamiltonian (\ref{ham}) by using the Green's function method,  and perform a mean-field approximation, thereby defining the condensate wavefunction $\psi_b^0\approx \big<\hat{\psi}^{\phantom{\dagger}}_b\big>  $, and each of the wavefunctions corresponding to the fermions, $\psi^n_{f}\approx\big<\hat{\psi}^{n}_f\big> $. We consider the quasi-one dimensional limit of a strongly elongated trap  with the transverse trapping frequency $\omega_\perp$, which leads to rescaling of the one-dimensional interaction coefficients $g_{bb}=2\hbar\omega_\perp a_{bb}$ and $g_{bf}=2\hbar\omega_\perp a_{bf}$. Thus, we arrive to a set of $N_f+1$ coupled mean-field equations, $N_f$ equations corresponding to each of the fermions and one for the BEC wavefunction \cite{salerno:063602}: 
\begin{eqnarray}
\label{mfeqm2}
\ii\hbar\partial_t \psi_f^n  &=&(\hat{H}_{f}+u_{bf} \rho_{b}^0)\psi^n_{f},
\\
\label{mfeqm1}
\ii\hbar\partial_t \psi_b^0  &=&(\hat{H}_{b}+u_{bb}\rho_{b}^0+u_{bf}\rho_{f}^0  )\psi^0_{b},
\end{eqnarray}
where,
$
\rho_{f}^0  =\alpha \sum_{n=1}^{N_{f}}|\psi^n_{f}  |^2 $,
$\alpha$ is a size parameter of the cloud,
and $\rho_{b}^0 =|\psi^0_{b}  |^2$.
The number of bosons (fermions) in the system is given by:  $\quad N_{b,f}=\int_{\mathbb{R}}\mathrm{d}x\rho^0_{b,f}$. By introducing the scaling units of time, $2/\omega_b$, length, $\sqrt{\hbar/ (m_b\omega_b)}$, and energy, $\hbar \omega_b$, the model can be re-written in the following dimensionless form:
\begin{eqnarray}
\label{ZDmfeqm2}
\ii\frac{m_f}{m_b}\partial_t \psi_f^n
+\partial^2_x\psi^n_{f}-V_{f}\psi^n_{f} +u_0|\psi^0_{b}  |^2\psi^n_{f}  =0,
\\
\label{ZDmfeqm1}
\ii\partial_t \psi_b^0+\partial^2_x\psi^0_{b}-V_{b}\psi^0_{b}+\sigma |\psi^0_{b}  |^2\psi^0_{b}
+u_1\rho_{f}^0 \psi^0_{b}=0,
\end{eqnarray}
where the bosonic wavefunction and fermionic density are rescaled as $\psi^0_{b}\rightarrow\psi^0_{b}[\hbar\omega_b/(2 |u_{bb}| )]^{1/2}$ and $\rho_f^0\rightarrow\rho_f^0 \hbar\omega_f/(2|u_{bb}|)$, respectively,
$u_0=-u_{b f } m_f/(|u_{b b}|m_b)$, $u_1=- u_{bf}\omega_f/(|u_{bb}|\omega_b)$, $\sigma=-\sgn(u_{b b})$. The double-well potential for bosons or fermions, $V_{\xi}=-m_\xi\kappa_{\xi}(|x|-x_0)^2/{m_b}$, is parametrized by the ratio of the trapping strengths, $\kappa_{\xi}=({m^{\phantom{2}}_{\xi}\omega^2_{\xi}})/({m^{\phantom{2}}_b\omega^2_b})$ that determines the spatial scale of the effective trapping potentials experienced by the different species of atoms.

\section{Coupled-mode theory}

Numerical solutions of the model equations (\ref{ZDmfeqm2}) are trackable when the number of fermions is small \cite{salerno:063602}. However, the current state-of-the art experiments  \cite{ospelkaus:020401} suggest that  the system should be considered in the limit of large number of atoms, both for bosonic and fermionic component, and the number of degenerate fermions can be of the same order of magnitude as the number of bosons. We consider the density of  fermions, $\rho_{f}^0$,  taking its stationary limit near the condensation temperature of the bosons, with the dynamics of the fermionic cloud occuring on a {\em much larger time scale} than that of the bosonic cloud.  In this limit, the fermions can be described in a first approximation by a filled Fermi sea, where the Fermi points $\pm k_F$ are invariant upon the interaction, and Luttinger's theorem  holds for the fermionic component \cite{PhysRev.119.1153,PhysRevLett.79.1106}. We assume that this quasistatic distribution of the fermionic density takes the form:
\begin{equation}\label{fdensity}
\rho_{f}^0  \approx\rho_{f}^0(x,0)\approx \frac{ r_{f b} N_b}{2} \delta\left(\frac{|x|-x_0}{\alpha}\right),
\end{equation}
where $r_{f b}=N_f/N_b$ is the fraction of fermions  in the system, which corresponds to
the fermions spatially localized at the minimum of each well (as shown in Fig. \ref{schematics}). The size of the localization region is governed by the parameter $\alpha\approx \sqrt{m_b/(m_f\kappa_{f})}$, with larger values of $\alpha$ corresponding to greater spatial extent of the fermionic cloud. The choice of density function (\ref{fdensity}) also ensures that the number of fermions in each well is conserved.

The assumption of a quasistatic fermionic density distribution introduced above neglects the dynamics of individual atoms given by (\ref{ZDmfeqm2}), as well as the feedback of the bosons on the fermions, while imposing a boundary condition on the total density. However, this approximation allows for analytical treatment of the model equation  (\ref{ZDmfeqm1}) governing the behaviour of bosons.  

In accordance with the coupled-mode theory for Bose-Einstein condensates trapped in a double-well potential, developed in \cite{PhysRevA.61.031601}, we assume that the main contribution to the bosonic condensate wavefunction comes from the two lowest {\em nonlinear modes} of the double-well potential,  corresponding to the symmetric ground state, $\Phi_0(x)$, and anti-symmetric first excited state, $\Phi_1(x)$:
\begin{equation}\label{modes}
\psi_b^0  =
\sum_{j=0}^1\Phi_j(x)B_j(t)\exp({-\ii \mu_j t+\sigma C_j n_j t})/{\sqrt{n_j}},
\end{equation}
where $B_j(t)$ is the time-dependent amplitude of the relevant state ($j=0,1$), $\mu_j$ is its energy,  $n_j=\int_{\mathbb{R}}\mathrm{d }x\;\Phi_j^2$  and $C_j=\int_{\mathbb{R}}\mathrm{d }x\;\Phi_j^4/n^2_j$. The nonlinear modes $\Phi_j$ obey the following stationary equations derived from Eq.  (\ref{ZDmfeqm1}):
\begin{equation}
\label{gpeXmodfer}
\frac{d^2 \Phi_j}{d x^2}+\mu_j\Phi_j-(|x|-x_0)^2\Phi_j+\sigma\Phi_j^3
+u_1\rho_f^0(x)\Phi_j=0,
\end{equation}
The equation (\ref{gpeXmodfer}) is equivalent to a standard mean field Gross-Pitaevskii equation for a BEC with a position dependent potential term modified by the nonlinear interaction with the fermionic cloud.   

Substituting the ansatz (\ref{modes}) into Eq. (\ref{ZDmfeqm1}) and  using Eq.  (\ref{gpeXmodfer}), we recover the system of  two coupled equations for the mode amplitudes \cite{PhysRevA.61.031601}:
\begin{eqnarray}
\ii \frac{d B_0}{d t}=\sig C_0|B_0|^2 B_0+\sig C_{0 1}\left( 2 |B_1|^2 B_0+B_0^*B_1^2\ee^{-\ii\Omega t}\right),
\nonumber
\\
\label{unsys1}
\ii \frac{d B_1}{d t}=\sig C_1|B_1|^2 B_1+\sig C_{0 1}\left( 2 |B_0|^2 B_1+B_1^*B_0^2\ee^{-\ii\Omega t}\right),
\end{eqnarray}
where: $C_{0 1}=\int_{\mathbb{R}}\Phi_0^2\Phi_1^2\mathrm{d }x/{(n_0 n_1)}$, and 
$\Omega=2(\be_1-\be_0)+2\sig(C_1 n_1 - C_0 n_0)$. This system of non-liner coupled equations describes the dynamical population exchange between the lowest energy states of the  condensate cloud in a double-well potential, and provides an alternative to a coupled mode description in the basis of ground states of the two separate wells. The advantage of the model (\ref{unsys1}) is in its ability to treat the system for {\em any} separation between the wells, providing an accurate picture of the population dynamics in the situation when the tunneling between the wells is strong. As shown in \cite{PhysRevA.61.031601}, the dynamical system (\ref{unsys1})  admits phase-locked solutions corresponding to macroscopically self trapped states, characterised  by the arrest of tunneling and formation of a stationary atomic density distribution which is nonzero in each well but strongly unbalanced. 

In order to analyze possible regimes of the self-trapping in the presence of localized fermionic impurities, we will set the physical parameters of the system to those of a $^{40}$K-$^{87}$Rb quantum degenerate mixture \cite{ferlaino:040702, ospelkaus:020401, albiez:010402,  PhysRevLett.89.150403, PhysRevLett.87.080403}.  We study the system where the rescaled one-dimensional s-wave inter-species interaction parameter is attractive, $u_{bf}\approx-234\, \mathrm{a}_0$ for  $^{40}_{\phantom{40}}\textrm{K}$ -$\,^{87}_{\phantom{87}}\textrm{Rb}$ and  the  analogous repulsive case, with $u_{bf}\approx+234\,\mathrm{a}_0$, where $\mathrm{a}_0$ is the Bohr radius. The boson-boson scattering parameter is $u_{bb}\approx98.98\,\mathrm{a}_0$ \cite{ospelkaus:020401}. The double-well angular frequency for $\;^{87}_{\phantom{87}}\textrm{Rb}$ is $\omega_{b}\approx 2\pi\times263$ Hz (see \cite{albiez:010402}) and  $\omega_{f}=\omega_{b}\sqrt{m_b/(m_f\kappa_{f})}\approx \omega_b/\alpha$.  In a realistic experimental situation $\kappa_{f}\approx1$ \cite{zaccanti:0606757}. In the following section, we examine the difference in the self-trapping scenario for bosons caused by the different types of inter-species interactions.

\section{Self-trapping regimes}

In order to determine the parameter space where MQST occurs, we use the ansatz, $B_j(t)=\sqrt{n_j(t)}\exp[-\ii\phi_j(t)]$, which allows us to rewrite the system (\ref{unsys1})  in terms of the the difference in relative populations of the modes $\Delta=n_1-n_0$ and the relative phase shift,  $\varphi=2 (\phi_0-\phi_1)-\Omega t$ \cite{PhysRevA.61.031601}:
\begin{eqnarray}\label{delta_phi}
&&\frac{d\Delta}{d t}=\sigma C_{01}(n^2-\Delta^2)\sin\varphi,
\nonumber
\\
&&\frac{d\varphi}{d t}=-\delta+\sigma(C_0+C_1)\Delta-2\sigma C_{01}(2+\cos\varphi)\Delta,
\end{eqnarray}
where $\delta=2(\mu_1-\mu_0)+\sigma[(n-2 n_0)C_0-(n-2 n_1)C_1]$, and $n=n_0+n_1=const$. The self-trapped states correspond to the regime where the relative phase shift is fixed to an integer of $2\pi$, $\varphi=2\pi m$. The difference in the population for the MQST states is given by the self-trapping parameter:
\begin{equation}
\label{TP}
\Delta_{01}\approx\frac{\delta}{\sigma (C_0+C_1-6 C_{01})},\;-n\leq\Delta_{01}\leq n.
\end{equation} 
The limiting cases correspond to the situation when only symmetric or antisymmetric mode is populated, resulting in the equal number of atoms in each well. The ultimate self-trapped state, when all the atoms are localized in one well, corresponds to $\Delta_{01}=0$.

The value of the self-trapping parameter $\Delta_{01}$ is strongly influenced by the mode coupling strengths $C_{i j}$ and the effective energies $\mu_j$ of the nonlinear modes. In our model these can be determined semi-analytically by means of a variational approach. This approach  employs an ansatz for the macroscopic wavefunction of the condensate, $\Phi$, in the form of a linear combination of the symmetric $\Phi_0$, and anti-symmetric $\Phi_1$ eigenstates of the double-well potential: $\Phi(x,x_0)=\Phi_0(x,x_0)+\Phi_1(x,x_0)$, where:
\begin{equation}
\label{trial}
\Phi_{0,1}(x)=A_{0,1}
\left(
\ee^{-\frac{(x-x_0)^2}{2 a_{0,1}^2}}
\pm
\ee^{-\frac{(x+x_0)^2}{2 a_{0,1}^2}}
 \right),
\end{equation}
and the amplitudes, $A_j$, and widths, $a_j$, are the variational parameters to be determined for each $\mu_j$.  The form of this ansatz reflects the existence of the first two lowest energy eigenstates, with symmetric and anti-symmetric spatial profiles and maxima localized  in the minima of the double-well potential. The details of the calculations are presented in the Appendix. Hereafter we present the results of the variational analysis.

\begin{figure}[ htbp!] 
   \centering
   \includegraphics[width=\columnwidth,keepaspectratio]{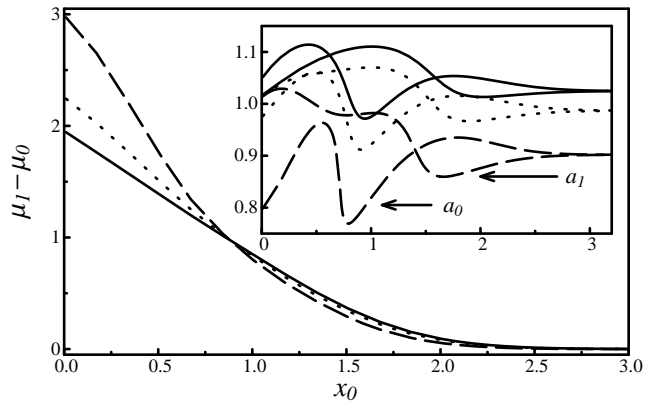} 
   \caption{The difference in energies, $\mu_1-\mu_0$, as a function of the double-well separation $x_0$. Inset: variation of the widths of the condensate wavefunctions, $a_{0,1}$, vs. ${x_0}$.  Parameters are: $r_{fb}=0$ (solid line), $0.2$ (dotted), $0.625$ (dashed), and $\alpha=0.77$.}
   \label{fig:ENneg}
\end{figure}

\subsection{Inter-species attraction}

We begin by analyzing the difference in energies, $\mu_1-\mu_0$, as a function of the double-well separation $x_0$ (see Fig.\ref{fig:ENneg}), in the case of $u_{bf}<0$, i.e. attraction between the fermionic impurities and BEC. In this case the growth in the concentration of fermions, $r_{fb}$, leads to the growing energy splitting between anti-symmetric and symmetric state, as compared to the pure BEC. The growth of the energy difference leads to the stronger localization of  the condensate wavefunction in each well (see the inset in Fig.\ref{fig:ENneg}), which in turn results in the suppression of tunneling  and the onset of the MQST at smaller well separations, compared to a pure BEC case. 

\begin{figure}[htbp!] 
   \centering
   \includegraphics[width=\columnwidth, keepaspectratio]{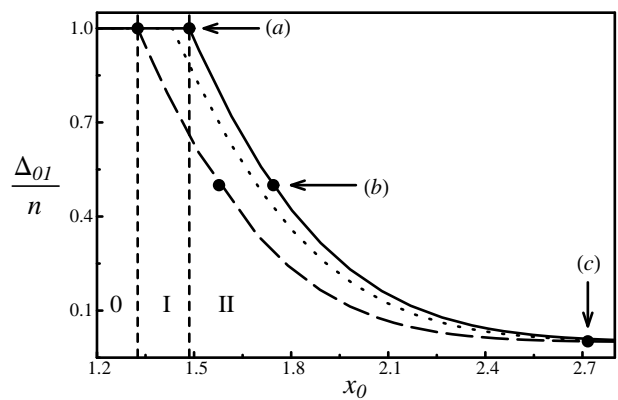} 
   \caption{The macroscopic quantum self-trapping parameter $\Delta_{01}/n$ as a function of the well separation, ${x_0}$, for attractive inter-species interaction. Parameters are: $r_{fb}=0$ (solid line), $0.2$ (dotted), $0.625$ (dashed), and $\alpha=0.77$. The roman numerals stand for regions with different symmetry properties of the BEC wavefunction (see text). } 
   \label{fig:DTneg}
\end{figure}

\begin{figure}[ht] 
    \centering
    \includegraphics[width=\columnwidth,keepaspectratio]{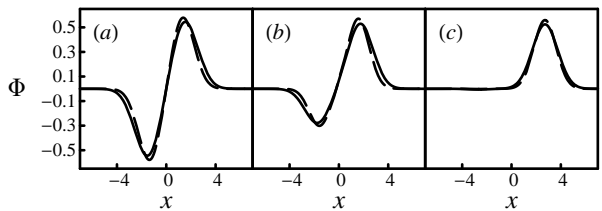} 
   \caption{The macroscopic wavefunction, $\Phi(x,x_0)$, for $r_{f b}=0$ (solid line) and $0.625$ (dashed) with attractive inter-species interaction and $x_0$ corresponding to the marked points in Fig.\ref{fig:DTneg}.}
   \label{fig:WFneg}
\end{figure}

The behaviour of the population imbalance, $\Delta_{01}(x_0)$, is shown in Fig.\ref{fig:DTneg}. In the region marked $0$ there is no self-trapping effect. In contrast, for region $I$, self-trapping occurs at smaller separation than the pure bosonic system, the latter restricted to the region $II$. As we increase the separation the system moves deeper into the self-trapped regime, where the  macroscopic wavefunction is completely localized in one of the two wells, and both the symmetric and antisymmetric mode are equally populated. 
Spatial profiles of the BEC wavefunction for the different regions in the parameter space are shown in Fig.\ref{fig:WFneg}. We can see that, although the attractive inter-species interaction leads to \emph{fermion induced narrowing } of the wavefunction peaks, the effect is weak even for large concentrations of fermions. Nevertheless, the difference in the strength of the coupling between the nonlinear modes of the system produced by the small variations in the spatial localization may lead to a significant fermion-induced extension of the self-trapping regime.  

The data from the experiments on the Bose-Josephson junction \cite{albiez:010402} provide a reference for experimentally feasible well separation. Provided that the Bose-Fermi mixture can be achieved with a strong inter-species interaction (i.e. either a wide spread of fermionic impurities, given by $\alpha$, or $|u_{bf}|/a_0 \gg 1$), the detection of the MSQT regime in the mixture is possible within the range of $x_0$ to $6.65$ and $10.15$, which corresponds to the experimentally achieved range of well separations from 4.4 $\mu$m to 6.7 $\mu$m.   We note that the strength of the nonlinear interaction between the BEC and fermionic impurities can be adjusted by modifying either the inter-species scattering length or the trapping frequencies for the two species ($\omega_b\approx\alpha\omega_f $).

\subsection{Inter-species repulsion}

\begin{figure}[htbp!]
\centering
   \includegraphics[width=\columnwidth, keepaspectratio]{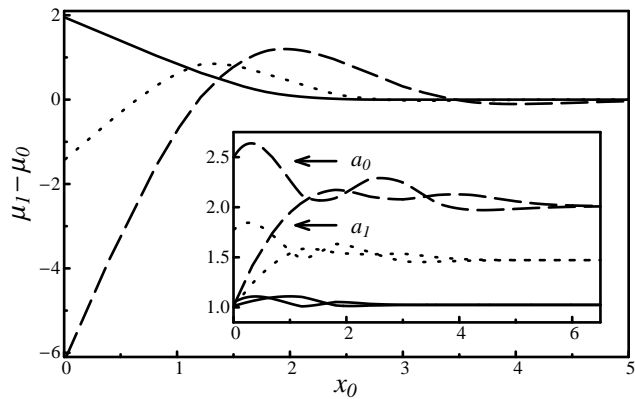} 
   \caption{The difference in energies, $\mu_1-\mu_0$, as a function of the double-well separation $x_0$ for repulsive inter-species interaction. Inset: the variation of the widths, $a_{0,1}$ vs. ${x_0}$. Parameters are: $r_{fb}=0$ (solid line), $0.2$ (dotted), $0.625$ (dashed), and $\alpha=12.2$.}
   \label{fig:ENpos} 
\end{figure}

The energy splitting between the ground and the first excited state of  BEC in the double-well potential in the case of repulsive inter-species interaction, $u_{fb}>0$, has a very different behavior compared to the attractive case. In the limit of small separations, the symmetric state is the lowest energy state in both pure BEC and Bose-Fermi mixture. In the pure BEC case the energy splitting monotonically tends to zero resulting in the energy degeneracy at large separations. However, the dependence of the energy splitting on the well separation in a Bose-Fermi mixture is non-monotonic, with an extremum value different from zero, see Fig.\ref{fig:ENpos}. For a certain range of well separations, namely in the region where the difference $\mu_1-\mu_0$ changes sign, the ground state of the bosonic component of the mixture becomes {\em antisymmetric}.

\begin{figure}[htbp!] 
   \centering
   \includegraphics[width=\columnwidth, keepaspectratio]{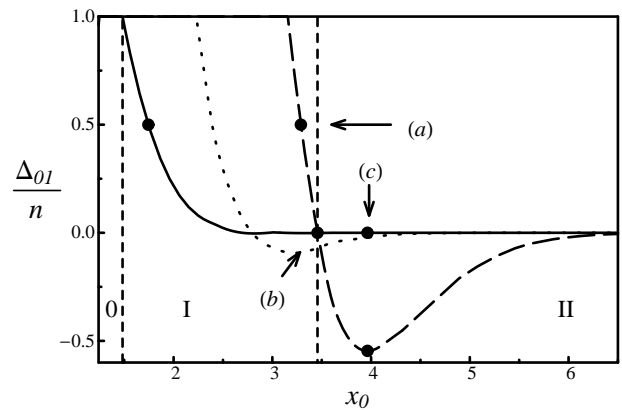} 
   \caption{The macroscopic quantum self-trapping parameter $\Delta_{01}/n$ as a function of the well separation,  ${x_0}$, for repulsive inter-species interaction. Parameters are: $r_{fb}=0$ (solid line), $0.2$ (dotted), $0.625$ (dashed), and $\alpha=12.2$. The roman numerals stand for regions with different symmetry properties of the BEC wavefunctions (see text).} 
   \label{fig:DTpos}
\end{figure}

\begin{figure}[htbp!] 
    \centering
    \includegraphics[width=\columnwidth, keepaspectratio]{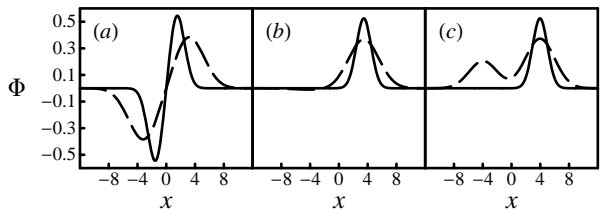} 
   \caption{The macroscopic wavefunction,
   $\Phi(x,x_0)$, for $r_{f b}=0$ (solid line) and $0.625$ (dashed) with repulsive inter-species interaction and $x_0$ corresponding to the marked points in Fig.\ref{fig:DTpos}.}
   \label{fig:WFpos}
\end{figure}

The dependence of the self-trapping parameter $\Delta_{01}$ on the well separation is shown in Fig. \ref{fig:DTpos}. The MQST states exist in all regions except for the region $0$. The onset of the MQST effect in a Bose-Fermi mixture occurs at larger well separations compared to that in a pure BEC system (see region $I$). This is due to the fact that the presence of repulsive fermionic impurities leads to the effective broadening  of the BEC wavefunction [see Fig.  \ref{fig:DTpos} (inset) and Fig. \ref{fig:WFpos} (a)], which extends the regime of strong inter-well tunneling to larger well separations. At the well separation corresponding to the point $(b)$ all atoms are localized in one of the well, like in the pure BEC case. Beyond this point, in region $II$, new behavior emerges, where owing to the population inversion between the ground and the first excited state, the MQST states with nonzero population in both wells emerge once again. 
The bosonic wavefunction corresponding to the point $(c)$ in region $II$ of Fig. \ref{fig:DTpos} are shown in figure \ref{fig:WFpos}. It can be seen that the symmetry of the MQST state in this region is {\em inversed} compared to the pure bosonic case, namely the peaks of the wavefunction in the two wells are {\em in-phase}. This is in a sharp contrast to the "conventional" self-trapped out-of-phase states, which occur both in purely bosonic BEC and in the Bose-Fermi mixture [cf. Figs. \ref{fig:WFneg}(a) and \ref{fig:WFneg}(c)]. The \emph{fermion induced symmetry breaking} of the macroscopic wavefunction is due to the fermion-induced population inversion between the symmetric and the anti-symmetric mode, $n_1<n_0$. Population imbalance vanishes asymptotically at large well separations in region $II$, whereby the bosons once again occupy only one of the two wells.

\begin{figure}[htbp!] 
    \centering
    \includegraphics[width=\columnwidth, keepaspectratio]{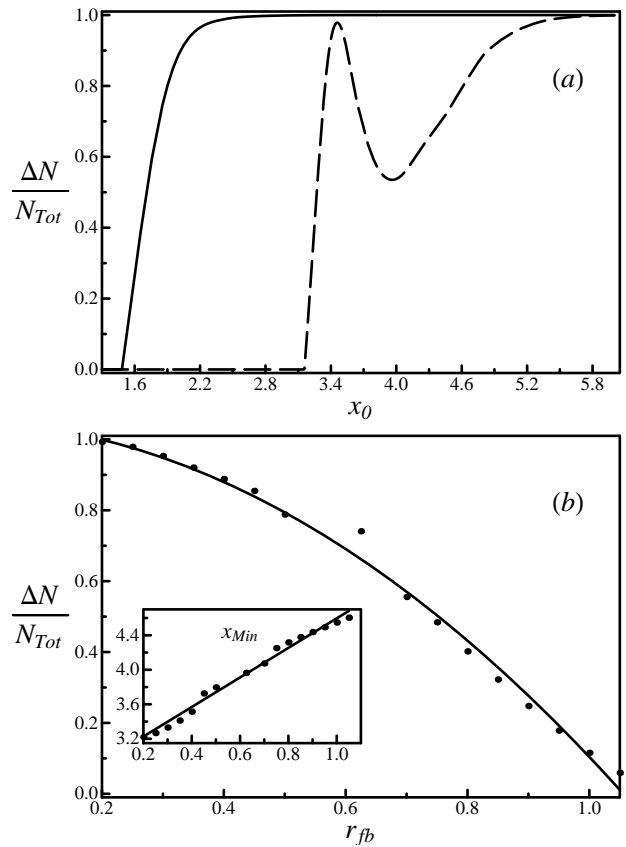} 
   \caption{(a) The relative population imbalance between the two potential wells in a self trapped state for repulsive fermionic component with the concentration $r_{fb}=0$ (solid line) and $0.625$ (dashed), as a function of well separation. (b) Minimum population imbalance in the symmetry-broken MQST state as a function of fermionic concentration. Solid line is the quadratic fit $\Delta N/N_{Tot}\approx 1.05 - 3.06 r_{fb} - 0.87 r^2_{fb}$. Inset: the value of well separation corresponding to the minimum imbalance as a function of $r_{fb}$. Solid line is the linear fit $x_{Min}\approx 2.88+ 1.71 r_{fb}$}
   \label{contrast}
\end{figure}

The highly nontrivial nature of the self-trapping in the case of repulsive interaction between the BEC and localized fermionic impurities can potentially be explored in an experiment with a double-well potential, similar to that presented in  \cite{albiez:010402}.  In an experiment, however, one would measure the difference in atom numbers between the BECs occupying two different wells of the potential, $\Delta N$, rather than the populations of two nonlinear modes $\Phi_{0,1}$. Possible results of such a measurement are shown in Fig. \ref{contrast}(a). It can be seen that, in contrast to the case of a pure BEC (solid line), the population imbalance, $\Delta N/N_{Tot}$, where $N_{Tot}$ is the total number of BEC atoms, is a nonmonotonic function of well separation [dashed line in Fig. \ref{contrast}(a)] approximately given by:
\begin{equation}
\frac{\Delta N}{N_{Tot}}=\left(\frac{1-\Delta_{01}^2}{1+\Delta_{01}
^2}\right)\frac{\erf(y)}{(1-\exp(-2 y))^{1/2}},\quad y=\frac{x_0}{a_1}
\end{equation}
At large concentration of fermions the population imbalance exhibits a sharp minimum at the certain value well of separation [see inset in Fig. \ref{contrast}(b)]. The minimum value of the imbalance tends to zero as $r_{fb} \to 1$ which means that, at large well separations, the self-trapping effect not only becomes less pronounced with the growth of the fermionic concentration, but may disappear entirely. We should note here that our assumption of the static fermionic density localized around the two minima of the potential wells becomes less reliable for $r_{fb}>1$.

\section{Conclusions}

We have analyzed the formation of self-trapped states in a BEC cloud mixed with degenerate fermions and confined in a one-dimensional double-well potential. Our analytical approach is reliable in the limit of large number of fermions and is based on the quasi-static approximation for the fermionic density, where the fermions are considered to form large impurities spatially localized around the minima of the two potential wells. The properties of the bosonic macroscopic wavefunction are analyzed by means of variational method, which allows us to comprehensively describe the spatial properties and symmetry of the self-trapped state. The self-trapping regimes in the Bose-Fermi mixture are predicted to be markedly different for repulsive and attractive inter-species interaction and highly sensitive to the fermion concentration. For the attractive interaction, the growth of the fermionic fraction has a marginal effect on the bosonic self-trapping, but nevertheless leads to the MQST at shorter well separations. This is due to the effective {\em suppression of tunneling} owing to the narrowing of the bosonic wavefunction. In the repulsive case, the growth in the fermionic fraction has a strong effect on the dynamics of bosons and pushes the onset of the MQST regime to larger well separations, as a result of the strongly {\em enhanced tunneling} due to the fermion-induced broadening of the BEC wavefunction.

Remarkably, the repulsive fermionic impurities can cause the population inversion between the ground and the first excited states of the BEC in a double-well potential and subsequent inversion of symmetry of the MQST state. Ultimate consequence of this inversion is the {\em suppression of self-trapping} at large well separations and a non-monotonic character of the population imbalance between the two wells as a function of well separation. The latter is a striking feature of the self-trapping effect in a Bose-Fermi mixture that can be detected experimentally.

Both the symmetry breaking and the suppression of self-trapping in the BEC cloud mixed with degenerate fermions signal the existence of the new regimes of the dynamics and switching of BECs in atomic waveguides and nonlinear interferometers, which can potentially be explored in experiments on an atom chip \cite{chip}. They are also expected to have profound consequence for the formation and dynamics of the self-trapped states in the Bose-Fermi mixtures loaded into periodic potentials. Beyond the mean field, these effects may have an implication for the onset of the superfluid to Mott insulator (MI) transition in a lattice potential \cite{nature.415.39, PhysRevLett.81.3108, PhysRevB.40.546}, leading, in the case of repulsive interaction, to the inhomogeneous suppression of the MI regime and phase separation. These phenomena require further investigation.

\begin{acknowledgments} 
This work was partly supported by the Australian Research Council (ARC) through the Discovery and Centre of Excellence schemes. M.G. and S.F.C.B. would like to acknowledge the hospitality of the Max-Planck Institut f\"ur Physik Komplexer Systeme. S.F.C.B. was supported by the CONACYT scholarship No. 167651. 
\end{acknowledgments}

\appendix*
\section{Variational solutions for $C_j$ and $C_{01}$.}
The dynamics of the system (\ref{delta_phi}) and the population imbalance characterising a self-trapped state (\ref{TP}) are determined by the interaction dependent coupling parameters, $C_{j}$, $C_{01}$, and the mode energies, $\mu_j$. To compute them, we use the variational method with the trial function given by Eq.(\ref{trial}). The mode energies are found from the following functional:
\begin{equation}
\label{funcL}
\mu_j=-2\max\left[\left.\int_{\mathbb{R}}\mathrm{d }x\;\lL[\Phi_j]\right|_{\mu_j=0}\right],
\end{equation}
where the Lagrangian density  $\lL[\Phi_j]$ corresponding to the equations of motion (\ref{gpeXmodfer}) for each nonlinear mode can be separated in three parts:
$
\lL=\lL_{k}+\lL_b+\lL_f
$,
The kinetic energy contribution ($\lL_{k}$), the bosonic  ($\lL_b$) and the fermionic ($\lL_f $) mean field contributions are:
\begin{eqnarray}
\lL_{k}
&=&-\frac{1}{2}\left(\frac{\mathrm{d}\Phi}{\mathrm{d} x} \right)^2,\nonumber
\\
\lL_b
&=&\frac{1}{2}\left[\mu -k(|x|-x_0)^2\right]\Phi^2-\frac{\sigma}{4}\Phi^4,
\nonumber
\\
\lL_f
&=&-\frac{c_1}{2}\rho_f^0(x)\Phi^2. 
\nonumber
\end{eqnarray} 
With the trial function (\ref{trial}),  and the fermionic density in the form (\ref{fdensity}) all the components of the Lagrangian density can be calculated analytically: 
\begin{eqnarray*}
\int_{\mathbb{R}}\mathrm{d }x\;\lL_f[\Phi_{0,1}]&=&-\frac{c_1r_{fb}\alpha A_{0,1}^2}{2}\left(1\pm w_{0,1}^2\right)^2,
\\
\int_{\mathbb{R}}\mathrm{d }x\;\lL_{k}[\Phi_{0,1}]&=&
\pm\frac{ 
A_{0,1}^2\sqrt{\pi}}{2 a_{0,1}^3}
      \left[
      \left( 
     2 x_0^2-a_{0,1}^2  
         \right)
        w_{0,1}
         \mp a_{0,1}^2 
         \right] ,\nonumber
\end{eqnarray*}
\begin{eqnarray*}
\int_{\mathbb{R}}\mathrm{d }x\;\lL_b[\Phi_0]&=&
\mu_0 a_0 A_0^2\sqrt{\pi}\left(1+w_{0}\right
)\\
&+&\frac{\sigma a_0 A_0^4\sqrt{\pi}}{2\sqrt{2}}
\left(
1
+3 w_{0}^2+4 w_{0}^{3/2}
\right)
\\
&+&\frac{k a_0 A_0^2 \sqrt{\pi}}{2}
\left[
4 x_0^2 \erf\left(\frac{x_0}{a_0}\right)-a_0^2-4 x_0^2
\right.
\\
&-&\left.\left(
a_0^2+2 x_0^2
\right)
w_{0}
+\frac{8 x_0 a_0}{\sqrt{\pi}}w_{0}\right],
\end{eqnarray*}
\begin{eqnarray*}
\int_{\mathbb{R}}\mathrm{d }x\;\lL_b[\Phi_1]&=&
\mu_1 a_1 A_1^2\sqrt{\pi}\left(1-w_1\right)\\
&+&\frac{\sigma a_1 A_1^4\sqrt{\pi}}{2\sqrt{2}}
\left(
1
+3w_1^2
-4w_1^{3/2}
\right)
\\
&+&\frac{k a_1 A_1^2 \sqrt{\pi}}{2}
\left[
4 x_0^2 \erf\left(\frac{x_0}{a_1}\right)
-a_1^2-4 x_0^2\right.\\
&+&\left.\left( 
a_1^2+2 x_0^2
\right)
w_1
\right],
\end{eqnarray*}
where, $w_{0,1}=\exp(-x_0^2/a_{0,1}^2)$ and  the square amplitudes, $A_{0,1}^2=1/[2 a_{0,1}\sqrt{\pi}(1\pm w_{0,1})]$. The coupling coefficients $C_0$, $C_1$, and $C_{01}$ can now be computed explicitly as a function of the widths $a_0$ and $a_1$:
\begin{eqnarray*}
&&C_0=\frac{1+3 w_0^2+4w_0^{3/2}}{2\sqrt{2 \pi} a_0\left(1+w_0\right)^2},
\\
&&C_1=\frac{1+3 w_1^2-4w_1^{3/2}}{2\sqrt{2 \pi} a_1\left(1-w_1\right)^2},
\\
&&C_{01}=\frac{1- 2w_0w_1-2w_0^{\gamma_0}w_1-2w_0w_1^{\gamma_1}+w_0^{2\gamma_0}w_1^{2\gamma_1}}{2\sqrt{\pi\left(a_0+a_1\right)}\left(1+w_0\right)\left(1-w_1\right)}
\end{eqnarray*}
where, $\gamma_{0,1}=a_{0,1}^2/(a_0^2+a_1^2)$.  The computation of all quantities follows from numerically minimizing the functional (\ref{funcL}) with all numerical constants fixed, and $a_{0,1}$ as variational parameters. In the limit of $r_{fb} \to 0$ our variational results are consistent with the numerical results for the pure BEC system presented in \cite{PhysRevA.61.031601}.

\end{document}